\documentclass[twocolumn]{revtex4}
\usepackage{amssymb}
\usepackage{latexsym}
\usepackage{epsfig}
\usepackage{color}
\usepackage{float}
\usepackage{graphicx}
\usepackage{epstopdf}

\begin{document}

\title{Inflation in the R\'{e}nyi cosmology}
\author{H. Moradpour$^1$\footnote{h.moradpour@riaam.ac.ir}, A. H. Ziaie$^1$\footnote{ah.ziaie@riaam.ac.ir}, V. B.
Bezerra$^{2}$\footnote{valdir@fisica.ufpb.br}, S. Ghaffari$^1$\footnote{sh.ghaffari@riaam.ac.ir}}
\address{$^1$ Research Institute for Astronomy and Astrophysics of Maragha (RIAAM), Maragha 55134-441, Iran\\
$^2$ Departamento de F\'{i}sica, Univeidade Federal da Para\'{i}ba, Caixa Postal 5008, CEP 58051-970, Jo\~{a}o Pessoa, PB, Brazil}

\begin{abstract}
The description of the accelerated phases for the current and primordial cosmos has been obtained in the framework of R\'{e}nyi cosmology. The results show that corrections to the Friedmann equations, originated from differences between R\'{e}nyi and Bekenstein entropies, can describe both the current accelerated and inflationary eras. The effects of these corrections can be appointed to a hypothetical fluid which is classically unstable. Although inflation can be obtained in R\'{e}nyi cosmology as the vacuum solution, the values found for the parameters of the hypothetical fluid which supports the current accelerated and primary inflationary cosmos differ from each other.
\end{abstract}

\maketitle

\section{Introduction}

Various generalized statistical mechanics have been proposed to
study the systems with long-range interactions
\cite{non0,non1,abe,nn1,pla,fon,nn2,nn3}. Since the Bekenstein
entropy is a non-extensive measure, and in fact, gravity is a
long-range interaction, physicists are motivated to consider various
generalized statistical mechanics for studying the cosmic evolution,
and in general, the gravitational phenomena
\cite{rn3,anp,Kom1,Kom2,Kom3,Kom,EPJC,epjcr,smm,Tavayef,non4,non5,non6,non7,non13,5,thdb,st,sara,sep,non2,non21,prd,Kom4,Kom5,epl}. In this regard, the Tsallis entropy is a well known entropy measure
\cite{non1,fon,abe} defined as

\begin{eqnarray}\label{reyn01}
&&S_T=\frac{1}{1-q}\sum_{i=1}^{W}(P_i^q-P_i),
\end{eqnarray}

\noindent for a system including $W$ discrete states. In this
definition, $q$ is an unknown constant evaluated by fitting with
experiments (observations) \cite{pla,non1,fon,abe,nn3,nn1}. The
R\'{e}nyi entropy is also another useful one-parameter generalized
entropy measure defined as \cite{non0}

\begin{eqnarray}\label{reyn01}
&&\mathcal{S}=\frac{1}{1-q}\ln\sum_{i=1}^{W} P_i^q.
\end{eqnarray}

Since the Bekenstein entropy is non-extensive, it has been proposed as a suitable candidate for the Tsallis entropy ($S_T=\frac{A}{4}$) \cite{non2,EPJC,prd}, a claim proved by Majhi \cite{5}. Combining $S_T=\frac{A}{4}$ with the $\delta\equiv1-q$ definition and also the above equations, one easily get \cite{non2,EPJC,prd}

\begin{eqnarray}\label{renyi}
\mathcal{S}=\frac{1}{\delta}\ln(1+\frac{\delta}{4}A),
\end{eqnarray}

\noindent as the R\'{e}nyi entropy content of the system whose boundary has an area $A=4\pi r^2$ located at radius $r$
\cite{non2,non21,epjcr,prd}. The generalized uncertainty principle (GUP) can also motivate physicists to consider the R\'{e}nyi entropy instead of the Bekenstein entropy \cite{mczg}.

In Ref. \cite{prd}, considering Eq.~(\ref{renyi}) as the entropy of
the apparent horizon of the flat FRW universe ($r=\frac{1}{H}$), and applying the Clausius
relation to the horizon, modified Friedmann equations has been
obtained in the R\'{e}nyi cosmology. Thereinafter, it has been shown that the obtained
additional terms, compared to the standard Friedmann equations, may
describe the current accelerated
universe. Thus, based on this study, the probable non-extensive
features of spacetime may be responsible for the current accelerated
universe, a result motivating us to study the possibility of
obtaining a primary inflationary era in the framework
of R\'{e}nyi cosmology obtained in Ref. \cite{prd}.

In the standard inflationary model, a quantum field (inflaton)
satisfies certain conditions, such as slow rolling, supports the
inflationary era \cite{roos}. There are also another models for
describing this era regardless of the inflaton
\cite{inf1,inf2,inf3,inf4}. In this paper, firstly, the power of
R\'{e}nyi cosmology in describing the current accelerated era in
the absence of the integration constant obtained in Ref.
\cite{prd} and also any dark energy fluid is investigated. We are
also interested in studying the possibility of describing
inflation in R\'{e}nyi cosmology without considering an inflaton field.
We present our analysis in the next section, and give a summary in
the third section. In this paper units have been set so that
$G=\hbar=c=k_B=1$, where $k_B$ denotes the Boltzmann constant and
dot displays the time derivative.

\section{Acceleration in The R\'{e}nyi cosmology}

Based on Ref. \cite{prd}, which attributes the R\'{e}nyi entropy instead
of the Bekenstein entropy to the apparent horizon of a flat FRW
universe, the Friedmann equations are modified as

\begin{eqnarray}\label{mfe}
&&\!\!\!\!\!\!\!\!H^2-\delta\pi\ln(\delta\pi+H^2)+C=\frac{8\pi}{3}\rho,\\
&&\!\!\!\!\!\!\!\!H^2+\frac{2}{3}\dot{H}-\delta\pi\ln(\delta\pi+H^2)-\frac{2\dot{H}}{3(\frac{H^2}{\delta\pi}+1)}+C=\frac{-8\pi}{3}p,\nonumber
\end{eqnarray}

\noindent where $C$ is a constant of integration, and ordinary
energy momentum conservation law is obeyed by the energy source
$T^{\mu}_{\ \nu}=diag(-\rho,p,p,p)$ (filling the background)
leading to \cite{prd}

\begin{eqnarray}\label{rasta12}
\dot{\rho}+3H(\rho+p)=0.
\end{eqnarray}

\subsection*{Current cosmos}

In Ref.~\cite{prd}, the authors did not consider the $C=0$ case, and
thus, in the limit $\delta\rightarrow0$, their model reduces to
the standard Friedmann equation in the presence of the
cosmological constant. It means that the $C\neq0$ case imposes the
problems of the cosmological constant to the model. Hence, we will
focus on the $C=0$ case in our study. In fact, choosing this case,
we run away from the phenomenological problems of the existence of
a constant within the field equations \cite{roos}. Now, whenever
$p=0$, Eq.~(\ref{rasta12}) leads to $\rho=\rho_0 a^{-3}$,
and thus, inserting this into Eq.~(\ref{mfe}) we arrive at

\begin{eqnarray}\label{mde}
&&H^2-\delta\pi\ln(\delta\pi+H^2)=\frac{8\pi}{3}\rho_0(1+z)^3,\\
&&H^2+\frac{2}{3}\dot{H}-\delta\pi\ln(\delta\pi+H^2)-\frac{2\dot{H}}{3(\frac{H^2}{\delta\pi}+1)}=0,\nonumber
\end{eqnarray}

\noindent where $\rho_0$ and $z$ denote the current value of the
matter density and redshift, respectively. The relation
$1+z=\frac{1}{a}$ has also been employed. Eq.~(\ref{mde}) leads to

\begin{eqnarray}\label{0}
\frac{\dot{H}}{\frac{\delta\pi}{H^2}+1}=-4\pi\rho_0(1+z)^3,
\end{eqnarray}

\noindent which clearly indicates that $\dot{H}=0$ at
$z\rightarrow-1$, and thus, we will face a de-Sitter
universe in this limit. For deceleration parameter $q$, one obtains
$q(z\rightarrow-1)\rightarrow-1$ because

\begin{eqnarray}\label{0}
q=-1-\frac{\dot{H}}{H^2}=-1+\frac{4\pi\rho_0(1+z)^3}{H^2}(\frac{\delta\pi}{H^2}+1).
\end{eqnarray}

\begin{figure}[ht]
\centering
\includegraphics[width=2.5in, height=2.5in]{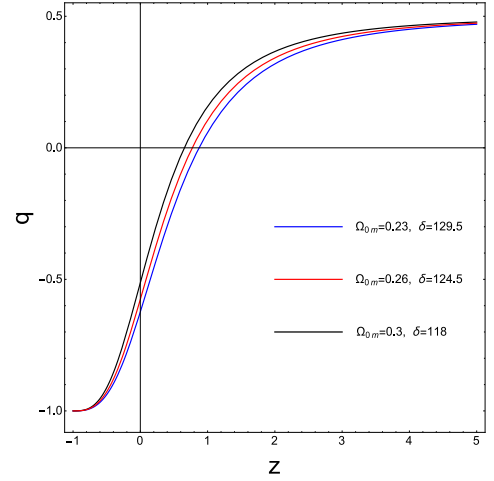}
\caption{\label{fig1} $q$ versus $z$ for $H(z=0)=67 s^{-1}$ and
some values of $\Omega_{0m}$. For the transition redshift $z_t$
(at which the universe expansion phase is changed from the
decelerated matter dominated phase to an accelerated expansion),
we have $z=0\cdot654$, $z=0\cdot771$ and $z=0\cdot873$ for
$\Omega_{0m}=0\cdot3$, $\Omega_{0m}=0\cdot26$ and
$\Omega_{0m}=0\cdot23$, respectively.}
\end{figure}

\noindent Moreover, using the definitions

\begin{eqnarray}\label{eff}
&&\rho_e=\frac{3\delta}{8}\ln(\delta\pi+H^2),\nonumber\\
&&p_e=-\rho_e-\frac{\dot{H}\delta}{4(H^2+\delta\pi)},
\end{eqnarray}

\noindent given in \cite{prd}, in order to rewrite
Eq.~(\ref{mde}), one can easily reach at \cite{prd}

\begin{eqnarray}\label{mfe0}
&&H^2=\frac{8\pi}{3}(\rho+\rho_e),\\
&&H^2+\frac{2}{3}\dot{H}=\frac{-8\pi}{3}p_e.\nonumber
\end{eqnarray}

\noindent In fact, in this way, we replaced a hypothetical fluid
with energy density and pressure $\rho_e$ and $p_e$, respectively,
by the modifications of the standard Friedmann equations
originated from the differences between the Bekenstein and
R\'{e}nyi entropies. The dimensionless density parameters are also
defined as

\begin{eqnarray}\label{omega}
\Omega_m=\frac{8\pi\rho}{3H^2},\ \Omega_e=\frac{8\pi\rho_e}{3H^2}.
\end{eqnarray}

\begin{figure}[ht]
\centering
\includegraphics[width=2.5in, height=2.5in]{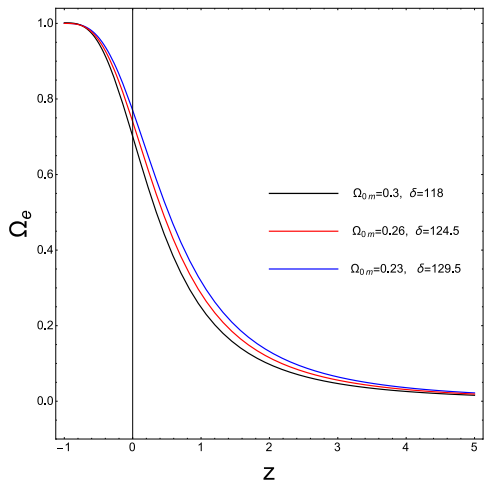}
\caption{\label{fig2} $\Omega_e$ versus $z$ for $H(z=0)=67 s^{-1}$ and some values of $\Omega_{0m}$.}
\end{figure}

\noindent The classical stability of the effective
source~(\ref{eff}) is evaluated by determining the sign of the
square of sound speed given by

\begin{equation}\label{ssp}
v_{s}^2=\frac{dp_e}{d\rho_e}=\frac{\dot{p}_e}{\dot{\rho}_e}=w_e+\dot{w}_e(\frac{\rho_e}{\dot{\rho}_e}),
\end{equation}

\begin{figure}[ht]
\centering
\includegraphics[width=2.5in, height=2.5in]{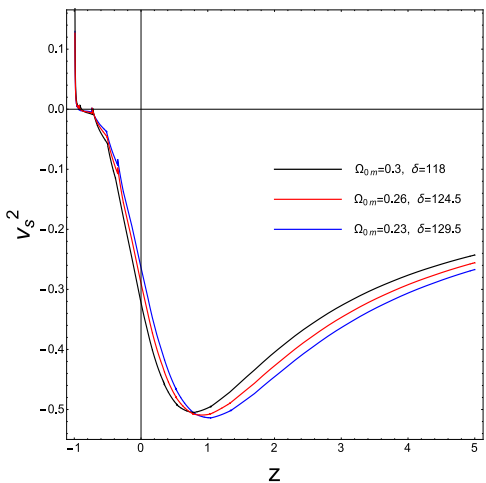}
\caption{\label{fig3} $v_{s}^2$ versus $z$ for $H(z=0)=67 s^{-1}$
and some values of $\Omega_{0m}$. Since we do not have always
$v_{s}^2>0$, the model is not classically stable.}
\end{figure}
\begin{figure}[ht]
\centering
\includegraphics[width=2.5in, height=2.5in]{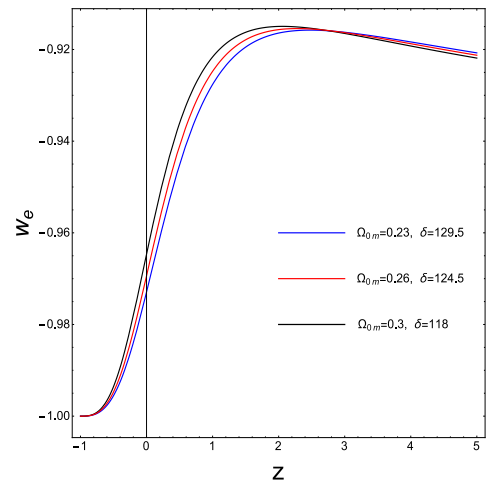}
\caption{\label{fig4} $w_e$ versus $z$ for $H(z=0)=67 s^{-1}$ and
some values of $\Omega_{0m}$.}
\end{figure}


\noindent where $w_e=\frac{p_e}{\rho_e}$ denotes the state
parameter of the effective source~(\ref{eff}). In
Figs.~(\ref{fig1})-(\ref{fig4}), the parameters $q$, $\Omega_e$,
$v_{s}^2$ and $w_e$ have been plotted, respectively, for the
initial condition $H(z=0)=67 s^{-1}$ and various values of
$\Omega_m(z=0)\equiv\Omega_{0m}$. In this situation, although the
model is not always classically stable (the $v_{s}^2>0$ condition
is not always satisfied), numerical results show that the
acceptable behavior of $\Omega_e$, $q$ and $w_e$ can be obtained
for $\delta\gtrapprox118$, depending on the value of $\Omega_{0m}$.

As it has been obtained in Eq.~(\ref{0}), we have $\dot{H}=0$ at
$z=-1$ or equally $H(z=-1)=constant$. Moreover, we have $\rho_m=0$
at $z=-1$ meaning that proper values of $\delta$ should lead to
$\Omega_e(z=-1)=1$, a condition that has been used to find some
suitable values for $\delta$ in plotting Fig.~(\ref{fig2}). The
values of $\delta$ obtained here (for $C=0$) are almost $100$
times greater than those found in~\cite{prd} where $C\neq0$.

\subsection*{Inflation without inflaton}

An interesting query is modelling the inflation in the absence of
inflaton \cite{inf1,inf2,inf3,inf4}. In the language of the
Friedman equations~(\ref{mfe}), it is equivalent to ask whether
the non-extensive features of spacetime prompt the universe to
experience an inflationary expansion or not. In order to answer
this question, we write Eq.~(\ref{mfe}) in vacuum as
\begin{eqnarray}\label{mfe2}
&&H^2-\delta\pi\ln(\delta\pi+H^2)=0,\\
&&H^2+\frac{2\dot{H}}{3(\frac{\delta\pi}{H^2}+1)}-\delta\pi\ln(\delta\pi+H^2)=0,\nonumber
\end{eqnarray}

\noindent which are available at the same time, only if
$\dot{H}=0$. This result indicates that the probable non-extensive
features of spacetime has theoretically enough power to describe
the primary inflationary era. Such an era lasted almost
$10^{-34}s$ leading to $H_{inf}\approx10^{34} s^{-1}$ for the
Hubble parameter at this inflationary era \cite{roos}.
Inserting this result into the first line of the above equation,
one reaches $\delta\approx0\cdot31$. It is worthwhile to mention
here that this value of $\delta$ cannot lead to a suitable
behavior for the current universe which means that the value of
$\delta$ in the inflation era differs from that of the current
universe.
\section{Summary and Concluding Remarks}

Considering the framework of R\'{e}nyi cosmology, we showed that
the probable non-extensive features of gravity formulated by the
R\'{e}nyi formalism may model the current accelerated universe in
a classically unstable mode without employing an ambiguous fluid
called dark energy such as a cosmological constant. It has also been
obtained that the vacuum solution of R\'{e}nyi cosmology provides
a setting for the inflationary phase.

In one hand, gravity is
described as the curvature of spacetime at the classical level. On
the other, the generalized entropy formalisms such as that of
R\'{e}nyi are used due to the long-range nature of gravity (or
equivalently the long-range nature of the spacetime curvature).
Therefore, one may argue that if the spacetime, and in fact, its
constituents satisfy the R\'{e}nyi entropy bound instead of that
of Bekenstein, in agreement with GUP \cite{mczg}, then differences between the R\'{e}nyi and
Bekenstein entropies may be responsible for the description of the accelerated expansion of the
universe. It is worthwhile mentioning that the values of $\delta$
in the current and primordial accelerated eras differ from each other
in this model.

\section*{Acknowledgment}
The work of H. Moradpour has been supported financially by Research
Institute for Astronomy \& Astrophysics of Maragha (RIAAM). V. B. Bezerra is partially supported by Conselho Nacional de Desenvolvimento Cient\'ifico e Tecnol\'ogico
(CNPq-Brazil) through the research project number. $305835/2016-5$

\end{document}